# Revisiting the configurations of hydrogen impurities in SrTiO$_3$: Insights from first-principles local vibration mode calculations

Zenghua Cai[1,2,*] and Chunlan Ma[1,2]

[1]*Key Laboratory of Intelligent Optoelectronic Devices and Chips of Jiangsu Higher Education Institutions, School of Physical Science and Technology, Suzhou University of Science and Technology, Suzhou, 215009, China*

[2]*Advanced Technology Research Institute of Taihu Photon Center, School of Physical Science and Technology, Suzhou University of Science and Technology, Suzhou, 215009, China*

*Contact author: zhcai@usts.edu.cn

**Abstract**

The specific configurations of hydrogen impurities in SrTiO$_3$ (STO) are still ambiguous. In this study, we systematically investigate the configurations and vibrational properties of hydrogen impurities in cubic STO using first-principles local vibration mode calculations. Employing the appropriate hybrid exchange correlation functional with the fraction of exact exchange setting to 0.2, we revisit the interstitial hydrogen (H$_i$), H$_i$ complexes (2H$_i$), and various intrinsic cation vacancy complexes with H$_i$, including V$_{Sr}$-H$_i$, V$_{Sr}$-2H$_i$, V$_{Ti}$-H$_i$, and V$_{Ti}$-2H$_i$. Comparison of the computed vibrational frequencies with experimental infrared absorption bands reveals that H$_i$, with a frequency of 3277 cm$^{-1}$, is unlikely to account for the dominant absorption bands near 3500 cm$^{-1}$. Instead, strontium vacancy complexes with interstitial hydrogen (V$_{Sr}$-H$_i$ and V$_{Sr}$-2H$_i$) exhibit vibrational frequencies that align with the main absorption bands, whereas titanium vacancy complex with two interstitial hydrogen (V$_{Ti}$-2H$_i$) corresponds to additional absorption bands around 3300 cm$^{-1}$. These findings provide insights into the importance of hydrogen-related complexes in governing the electronic properties of STO, and meanwhile underscore the necessity of employing accurate exchange correlation functionals for reliable theoretical predictions of vibrational properties.

## I.  INTRODUCTION

Hydrogen is an abundant and ubiquitous element in semiconductors[1], particularly in oxide semiconductors[2, 3] such as the prototypical perovskite oxide SrTiO$_3$ (STO). As early as 1968, Wakim has found the existence of hydrogen in STO.[4] The corresponding OH absorption band was observed at 3490 cm$^{-1}$ as a single peak. In 1986, Weber et al. detected the OH absorption bands at about 3500 cm$^{-1}$ in STO (three peaks),[5] and a year later, Houde et al. observed the similar OH absorption bands.[6] These early studies are primarily based on the infrared spectroscopic measurement. In 1994, Klauer et al. investigated the incorporation of hydrogen in STO using polarized Raman scattering.[7] They found a primary OH absorption band at 3496 cm$^{-1}$ (single peak) along with three additional, strong, high frequency absorption bands. In 2011, Tarun et al. reported two new absorption bands at 3355 and 3384 cm$^{-1}$ (two peaks) in addition to the three peaks around 3500 cm$^{-1}$.[8] So far the incorporation of hydrogen in STO mainly manifests in a single peak below 3500 cm$^{-1}$, three peaks above 3500 cm$^{-1}$, and two peaks around 3300 cm$^{-1}$. Many reports point out that the three peaks are essentially consistent with the single peak, as the former typically appear at cryogenic temperatures (about 4.2~14 K[5,6]), while the latter is generally observed at room temperature. The reason is that as STO transitions from cubic to the tetragonal phase when the temperature drops below 105 K, its symmetry decreases, leading to the splitting of the absorption band from a single peak to three peaks. In general, distinct OH absorption bands represent different configurations of hydrogen impurities, indicating that hydrogen can exist in STO in multiple forms.

Hydrogen can profoundly affect the electronic properties of semiconductors.[1] Therefore, the incorporation of hydrogen can significantly influence the electronic properties of STO. For example, Peacock et al. found that interstitial hydrogen acts as a shallow donor[9], contributing to the n-type conductivity.[10] Winczewski et al. reported that interstitial hydrogen induces significant polarization in STO.[11] Jalan and Iwazaki et al. observed that hydrogen can make the oxygen-deficient STO return into the insulating state from the conductive state.[12,13] Lin and Angelo et al. discovered that hydrogen can induce the (001) surface metallization of STO.[14,15] Furthermore, Ito found that hydrogen can result in the localized electron behavior in hydrogen-irradiated STO.[16]

In order to understand the influence of hydrogen impurities (doping) on the electronic properties of STO, many researchers aim to identify their specific configurations, i.e., the type and structure, based on the absorption bands. In experiment, Weber and Houde et al. did not specify the type of hydrogen doping[5,6], while they investigated the position of hydrogen. Weber et al. reported that the stretching vibration of O-H bond occurs along the oxygen-oxygen direction[5], suggesting that the hydrogen site is aligned accordingly. Houde et al. deduced that hydrogen resides on the faces of the cubic STO cell, between Sr and O atoms.[6] In 1994, Klauer et al. clearly identified the type of hydrogen doping as interstitial hydrogen ($H_i$), corresponding to the absorption bands with three peaks.[7] Their Raman data also showed that hydrogen is located between two next-nearest oxygen ions along the cubic axes. Tarun et al. attributed the two new absorption bands (3355 and 3384 cm$^{-1}$) to a strontium vacancy ($V_{Sr}$) passivated by two hydrogen atoms $V_{Sr}$-$2H_i$.[8]

In theory, Bork et al. calculated vibrational frequencies of $H_i$ and $2H_i$ (or $H_i$-$H_i$ complex).[17] The results are 3533 cm$^{-1}$ for $H_i$, and 3465 and 3438 cm$^{-1}$ for $2H_i$, thereby they attribute the single peak and two peaks observed in experiment to $H_i$ and $2H_i$, respectively. Limpijumnong et al. investigated the local vibration mode of several complex configurations of hydrogen impurities, including a strontium vacancy passivated by one ($V_{Sr}$-$H_i$) or two hydrogen atoms ($V_{Sr}$-$2H_i$), and a titanium vacancy ($V_{Ti}$) combined with different number of $H_i$ ($V_{Ti}$-$nH_i$).[18,19] The calculated vibrational frequency for $V_{Sr}$-$H_i$ is 3505 cm$^{-1}$. For $V_{Sr}$-$2H_i$, four configurations were considered, with type I and II among them exhibiting vibrational frequencies of 3523 and 3527 cm$^{-1}$, respectively. These configurations are thus ascribed to the experimentally observed three peaks.[20] Moreover, their calculations indicated that $V_{Ti}$-2$H_i$ is likely responsible for the two peaks around 3300 cm$^{-1}$. Varley et al. performed a systematic study on hydrogen impurities in STO through calculating the formation energies and vibrational frequencies for various configurations of hydrogen doping.[21] Their results suggest that $H_i$ is the source of the single peak, while $V_{Ti}$-2$H_i$ corresponds to the two peaks.

Despite extensive efforts, the specific configurations of hydrogen impurities in STO remain unclear. In experiment, only limited hypotheses regarding specific configurations have been proposed, owing to the inherent difficulty of identification in experiment. Theoretically, although numerous configurations have been suggested to explain the observed absorption bands, substantial divergence persists. For example, Bork and Varley et al. think that the main absorption bands (the single or three peaks) originate from $H_i$,[17, 21] whereas Limpijumnong et al. contend that $H_i$ cannot account for the main absorption bands.[19] Instead, they propose that the main absorption bands near 3500 cm$^{-1}$ are associated with the defect complexes consisting of $H_i$ and $V_{Sr}$. For the two peaks around 3300 cm$^{-1}$, both Limpijumnong and Varley et al. attribute them to $V_{Ti}$-2$H_i$,[19, 21] while Bork et al. assign them to 2$H_i$.[17] All these conflicting interpretations motivate us to revisit the configurations of hydrogen impurities in STO.

In this work, we performed a systematic study on the local vibration modes of hydrogen impurities in STO based on first-principles calculations. First, the formation energies of hydrogen doping are calculated so as to identify all possible configurations of hydrogen impurities. Subsequently, a numerical method is used to calculate the vibrational frequencies of these configurations. To ensure comprehensive coverage, all critical configurations reported in previous theoretical works are also reevaluated. At last, the calculated vibrational frequencies are meticulously compared with the experimentally observed absorption bands. This work revisits the specific configurations of hydrogen impurities in STO, and may provide some insights for understanding the behavior of hydrogen impurities in STO.

## II. THEORETICAL DETAILS

### A. Details of the calculations

All the first-principles calculations are performed based on density functional theory (DFT) as implemented in the VASP code[22]. Defect property simulations are carried out by using a 520 eV cutoff energy, a single Γ point (1×1×1 Monkhorst-Pack k-point mesh)[23] and a 3×3×3 supercell (135 atoms) based on the five-atom primitive cell of cubic $SrTiO_3$ (STO). Structural relaxations for the defect and vibrational frequency calculations are conducted based on the Perdew-Burke-Ernzerhof (PBE) exchange correlation functional within the generalized gradient approximation (GGA)[24] and the screened hybrid functional of the Heyd, Scuseria, and Ernzerhof[25] (HSE06), respectively, until the residual forces on all atoms are less than 0.01 eV/Å. HSE06 is used for all the total energy calculations with the fraction of exact exchange setting to 0.2 (α = 0.2) as adopted in recent work[26]. Defect properties are simulated by using the supercell model, following the procedure described in DASP code[27]. For analyzing local vibrational modes, which depend only on the defect configurations and corresponding charge states, defect properties under O-poor conditions are calculated as the candidate. Further details regarding the defect simulations are available in our previous work.[28]

### B. Details of the local vibration mode calculations

In order to calculate the frequency of the local vibration mode, we follow the approach used in the previous work.[29] First, a series of small displacements of the H atom are introduced along the direction of the O-H bond in both positive and negative directions. For each O-H bond of different configurations, 20 displacements are included with magnitudes up to ±30% of the bond length. Then, the total energies of different displacement configurations are calculated. The variation of total energy as a function of displacement distance is shown in Fig. 1(a). By fitting the results with an optimal fourth-degree polynomial, with the total energy at the equilibrium position set to zero, a potential-energy function can be obtained:

$$V(x) = \frac{k}{2}x^2 + \alpha x^3 + \beta x^4. \tag{1}$$

The harmonic frequency $\omega_0$ can be calculated through the coefficient of the quadratic term, i.e., $\omega_0 = \sqrt{k/\mu}$. $\mu$ is the reduced mass (0.9483 amu), which is defined as:

$$\frac{1}{\mu} = \frac{1}{m_H} + \frac{1}{m_O}, \tag{2}$$

where $m_H$ and $m_O$ are the masses of the H and O atoms, respectively. The coefficients $\alpha$ and $\beta$ are the high order terms that account for the anharmonic effects.

Once the potential energy function is obtained, we solve the following one-dimensional Schrödinger equation:

$$\left[-\frac{\hbar^2}{2\mu}\nabla^2 + V(x)\right]\psi(x) = E\psi(x). \tag{3}$$

The numerical solution of Eq. (3) with potential energy function given by Eq. (1) is obtained via the shooting method with 20000 grid points. The vibrational frequency $\omega$ is determined from the transition energy between the ground state ($E_0$) and the first excited state ($E_1$):

$$\hbar\omega = \hbar(\omega_0 + \Delta\omega) = E_1 - E_0, \tag{4}$$

where $\Delta\omega$ is the anharmonic contribution.

It is worth noting that the choice of exchange correlation functional can significantly influence the calculated vibrational frequency. The previous reports indicate that hybrid functional tends to overestimate the vibrational frequency.[30] In contrast, the local density approximation (LDA) tends to overestimate the interactions between H and next-nearest neighbor O atoms, thereby underestimating the vibrational frequency.[21] Recent findings indicate that hybrid functional (HSE06), with an exact exchange fraction of 0.2, can effectively describe the electronic properties of STO.[26] Hence, this functional and exchange setting are used in this work.

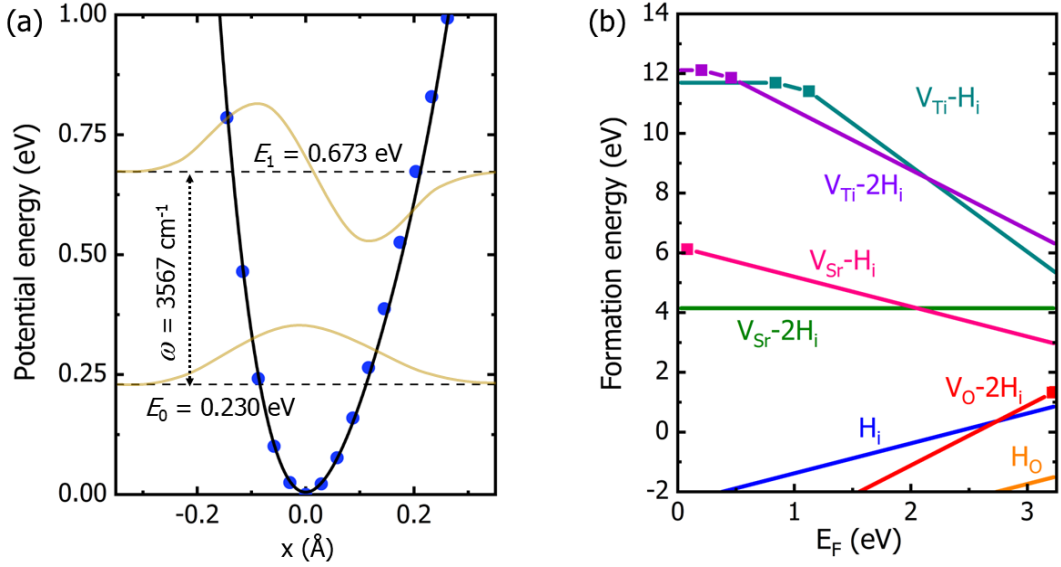

FIG. 1. (a) The calculated potential energy of $V_{Sr}$-$H_i$ as a function of displacement distance $x$ with respect to O-H equilibrium bond length. The blue dots represent the calculated potential energies relative to the energy of the equilibrium position. The black line is obtained through fitting the calculated potential energies based on Eq. (1). (b) The calculated formation energy of hydrogen doping as a function of the Fermi level ($E_F$) under O-poor condition. The chemical potentials of Sr, Ti, O and H are -0.83, -0.28, -5.34 and -0.65 eV, respectively.

## III. RESULTS AND DISCUSSION

### A．Formation energies of hydrogen-related point defects

The incorporation of hydrogen impurities in STO will introduce many types of point defects. In general, these point defects can be divided into two categories: interstitial and substitutional defects. For example, $H_i$ and hydrogen replacing oxygen ($H_O$) are the typical point defects of hydrogen doping in STO. Complex is another category of point defect, which is usually uncommon. However, due to its small atomic size, hydrogen readily forms complexes. In STO, the intrinsic vacancy defects ($V_X$, X=Sr, Ti, O) can easily trap several hydrogen atoms, forming defect complexes ($V_X$-$nH_i$, n=1, 2, 3, …) as reported previously.[18, 19, 21] For $V_X$-$nH_i$ complexes, previous works indicate that the absorption bands observed in experiment are mainly related with the complexes where n=1 or 2. Consequently, only these complexes are considered here. As hydrogen impurities typically induce n-type conductivity in STO, we concentrate on defect properties in n-type STO, where the Fermi level ($E_F$) is very close to the conduction band minimum. Fig. 1(b) illustrates the calculated formation energies of different hydrogen-related point defects under O-poor condition. As we can see, $H_O$ is a donor and exhibits the lowest formation energy, even lower than that of $H_i$. Meanwhile, $V_O$-$2H_i$ also acts as a donor, displaying the comparable formation energy with that of $H_i$ in n-type STO. Hence, $V_O$ occupied by hydrogen atoms serve as the significant donor defects, contributing to the n-type conductivity of STO under O-poor condition. This is consistent with the previous report.[13] For $V_{Sr}$-$nH_i$, they also have relatively low formation energies. $V_{Sr}$-$H_i$ is an acceptor and predominantly stable in the -1 charge state, while $V_{Sr}$-$2H_i$ is always stable in the neutral charge state. In contrast, $V_{Ti}$-$nH_i$ exhibit relatively high formation energies. In n-type STO, $V_{Ti}$-$H_i$ is stable in the -3 charge state, while $V_{Ti}$-$2H_i$ is stable in the -2 charge state. These results agree well with the previous report.[21] Through the analysis of different defect configurations, we find that O-H bond is present in most hydrogen-related point defects except for $H_O$ and $V_O$-$2H_i$. Moreover, for the point defects with O-H bonds, they are stable in various charge states in n-type STO. In order to facilitate the comparison with the experimentally observed absorption bands, only the specific configurations and charge states of these point defects are considered in the subsequent discussion.

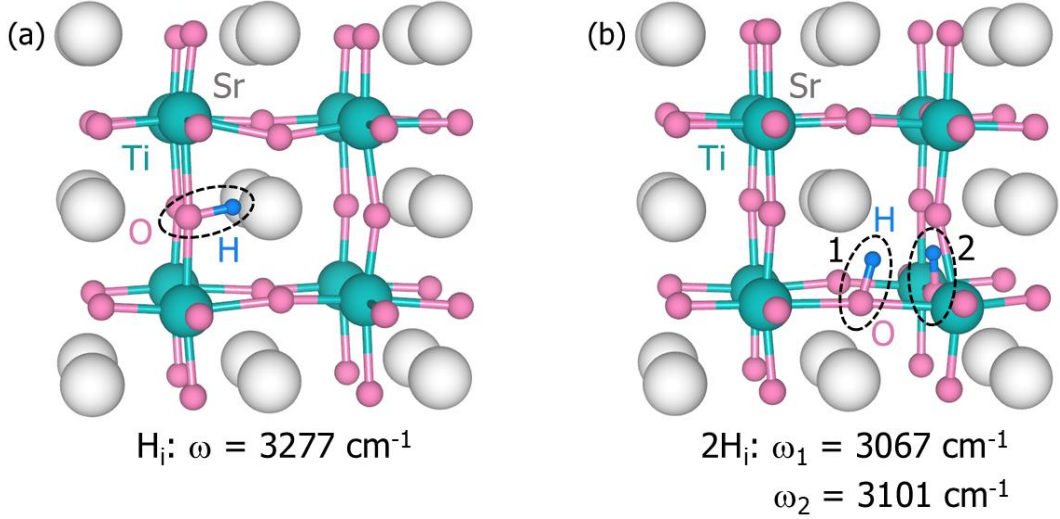

FIG. 2. (a) The configuration of $H_i$ in +1 charge state along with the calculated vibrational frequency for the corresponding O-H bond; (b) The configuration of $2H_i$ in neutral charge state along with the calculated vibrational frequencies for the corresponding O-H bonds. 1 and 2 represent the bond indices.

### B．Vibrational frequencies of interstitial hydrogen and the related complex

The interstitial hydrogen is the most simple and common configuration of hydrogen impurities. In oxide semiconductors, $H_i$ typically acts as a donor, contributing to the n-type conductivity.[10] For example, previous studies have found that $H_i$ prefers to be a donor and stay in the +1 charge state in n-type STO.[19, 21] Similarly, our results also indicate that $H_i$ is a donor in the +1 charge state as shown in Fig. 1(b). Fig. 2(a) displays the relaxed configuration of $H_i$ in +1 charge state. This configuration is consistent with previous reports.[19, 21] The O-H bond length of $H_i$ is 0.972 Å as listed in Table I, which is longer than 0.964 Å reported by Varley et al.[21] and shorter than 1.011 Å reported by Limpijumnong et al.[19] This difference primarily arises from the different exchange correlation functionals. Specifically, for the O-H bond, the standard HSE06 functional used by Varley et al. ($\alpha = 0.25$) may localize electrons more effectively, thus strengthening the O-H bond and leading to a relatively short bond length, whereas the LDA functional used by Limpijumnong et al. may delocalize electrons excessively, resulting in a weakened O-H bond and a longer bond length. As mentioned in Sec. II, the HSE06 ($\alpha = 0.2$) functional used in this work can effectively describe the electronic properties of STO. Therefore, a more accurate and moderate O-H bond length is obtained here. Based on the numerical method, the vibrational frequency of O-H bond for $H_i$ has been calculated. As listed in Table I, the harmonic frequency is 3538 cm$^{-1}$ and the anharmonic contribution is -261 cm$^{-1}$. As a result, the total vibrational frequency $\omega$ is 3277 cm$^{-1}$.

Besides the H$_i$, Bork et al. also considered a neutral H$_i$-related complex 2H$_i$ (or H$_i$-H$_i$).[17] For comprehensiveness, the neutral 2H$_i$ is also considered as shown in Fig. 2(b). In this configuration, two hydrogen atoms occupy distinct endpoints of a SrO$_6$ octahedral edge, forming two O-H bonds. Bond 1 (b$_1$) is 0.980 Å and bond 2 (b$_2$) is 0.979 Å. As listed in Table I, the calculated vibrational frequency for b$_1$ is 3067 cm$^{-1}$, with harmonic and anharmonic contributions of 3372 and -305 cm$^{-1}$, respectively. Meanwhile, the calculated vibrational frequency for b$_2$ is 3101 cm$^{-1}$, with harmonic and anharmonic contributions of 3393 and -292 cm$^{-1}$, respectively.

Table I. Vibrational properties for local vibration mode of the O-H bond in different configurations. $b_{index}$ is the bond index and $d_{O-H}$ is the bond length. $k/2$, $\alpha$ and $\beta$ are the second, third and fourth order coefficients of Eq. (1), in eV/Å$^n$, where $n$ is the order of the coefficients. $\omega_0$ and $\Delta\omega$ are the harmonic and anharmonic contributions to the vibrational frequency $\omega$, i.e., $\omega = \omega_0 + \Delta\omega$. All frequencies are in cm$^{-1}$.

| Type | $b_{index}$ | $d_{O-H}$ (Å) | $k/2$ | $\alpha$ | $\beta$ | $\omega_0$ | $\Delta\omega$ | $\omega$ |
|---|---|---|---|---|---|---|---|---|
| H$_i$ |   | 0.972 | 21.93 | -73.77 | 129.94 | 3538 | -261 | 3277 |
| 2H$_i$ | 1 | 0.980 | 19.91 | -70.74 | 123.72 | 3372 | -305 | 3067 |
|   | 2 | 0.979 | 20.16 | -71.11 | 125.62 | 3393 | -292 | 3101 |
| V$_{Sr}$-H$_i$ |   | 0.968 | 24.59 | -73.86 | 132.58 | 3747 | -180 | 3567 |
| V$_{Sr}$-2H$_{i1}$ | 1 | 0.968 | 24.47 | -74.03 | 132.19 | 3738 | -187 | 3551 |
|   | 2 | 0.968 | 24.41 | -74.35 | 133.25 | 3733 | -188 | 3545 |
| V$_{Sr}$-2H$_{i2}$ | 1 | 0.968 | 24.49 | -74.00 | 133.76 | 3739 | -179 | 3560 |
|   | 2 | 0.967 | 24.87 | -74.67 | 134.19 | 3768 | -179 | 3589 |
| V$_{Sr}$-2H$_{i3}$ | 1 | 0.969 | 24.34 | -73.59 | 133.48 | 3728 | -178 | 3550 |
|   | 2 | 0.968 | 24.71 | -74.34 | 134.20 | 3756 | -178 | 3578 |
| V$_{Ti}$-H$_i$ |   | 0.973 | 23.12 | -71.88 | 128.39 | 3633 | -199 | 3434 |
| V$_{Ti}$-2H$_{i1}$ | 1 | 0.976 | 21.61 | -71.64 | 127.01 | 3513 | -245 | 3268 |
|   | 2 | 0.978 | 21.03 | -70.57 | 125.01 | 3465 | -252 | 3213 |
| V$_{Ti}$-2H$_{i2}$ | 1 | 0.974 | 22.04 | -72.63 | 127.87 | 3547 | -246 | 3301 |
|   | 2 | 0.974 | 22.07 | -72.73 | 128.72 | 3548 | -241 | 3307 |

**C．Vibrational frequencies of strontium vacancy complexes with interstitial hydrogen**

The intrinsic strontium vacancy can trap several hydrogen atoms, forming V$_{Sr}$-nH$_i$ complexes, owing to the large atomic size of strontium atom. At present, V$_{Sr}$-H$_i$ and V$_{Sr}$-2H$_i$ represent the most common strontium vacancy complexes with interstitial hydrogen, and have been studied previously based on different exchange correlation functionals.[19, 21] In this work, we also focus on these two complexes. Fig. 3(a) shows the stable configuration of V$_{Sr}$-H$_i$ in -1 charge state after relaxation. In this configuration, the O-H bond is oriented towards the V$_{Sr}$, consistent with previous reports.[19, 21] The length of this O-H bond is 0.968 Å as listed in Table

I, which is nearly identical to 0.967 Å as indicated by Varley et al.[21] and shorter than 0.985 Å as reported by Limpijumnong et al.[19] The calculated vibrational frequency of this O-H bond is 3567 cm$^{-1}$, comprising a harmonic contribution 3747 cm$^{-1}$ and an anharmonic contribution -180 cm$^{-1}$.

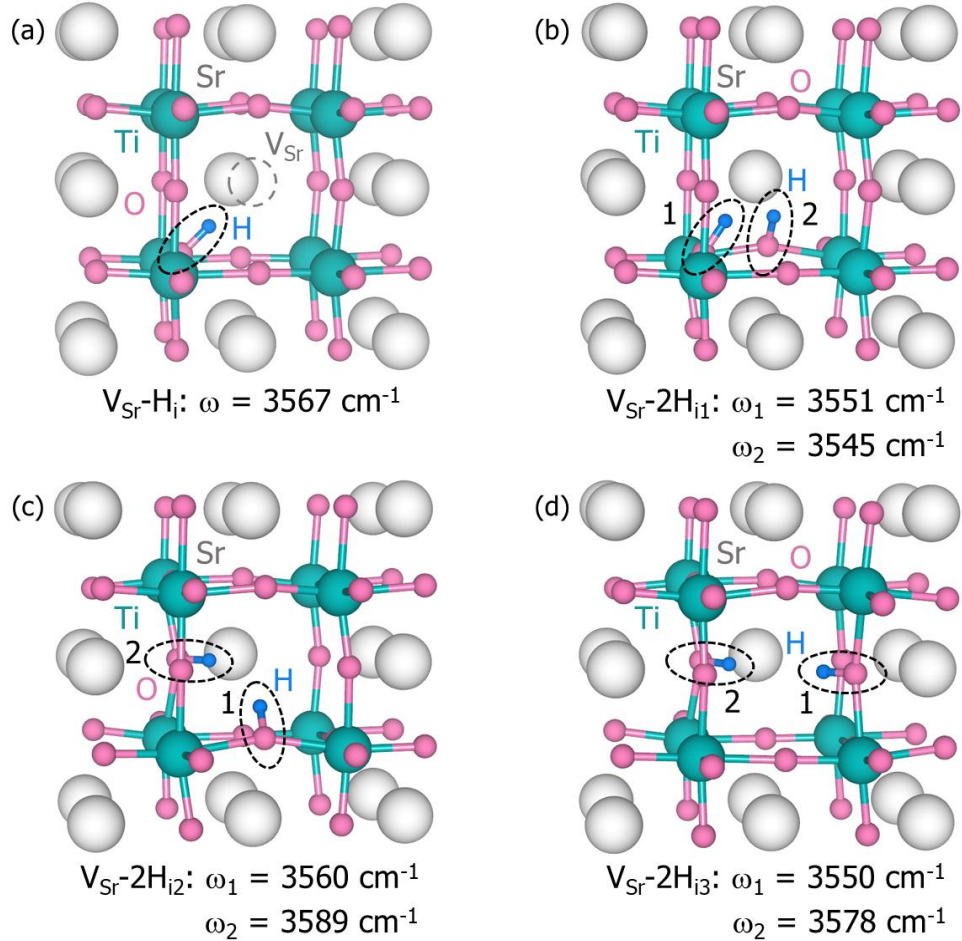

FIG. 3. (a) The configuration of $V_{Sr}$-$H_i$ in -1 charge state along with the calculated vibrational frequency for the corresponding O-H bond; (b), (c) and (d) are the different configurations of $V_{Sr}$-$2H_i$ in neutral charge state along with the calculated vibrational frequencies for the corresponding O-H bonds. 1 and 2 represent the bond indices.

For $V_{Sr}$-$2H_i$, our calculation results indicate that its configuration (see Fig. 3(b), denoted as $V_{Sr}$-$2H_{i1}$) exhibits nearly the same hydrogen positions as those in $2H_i$ complex as shown in Fig. 2(b). The bond lengths of $b_1$ and $b_2$ in $V_{Sr}$-$2H_{i1}$ are slightly shorter than those in $2H_i$ as listed in Table I. This configuration is also consistent with that reported by Varley et al., with the lengths of the corresponding O-H bonds being almost the same (0.968 Å vs. 0.967 Å).[21] Through the numerical calculations, the vibrational frequencies of the O-H bonds in this configuration have been obtained. The difference between the vibrational frequencies of b1 and b2 is minimal (only 6 cm$^{-1}$) as shown in Fig. 3(b). This small difference originates from that the harmonic and anharmonic contributions of these two O-H bonds are nearly identical as listed in Table I.

As mentioned in Sec. I, Limpijumnong et al. also reported two important types of $V_{Sr}$-$2H_i$ complexes, which possess configurations distinct from that of $V_{Sr}$-$2H_{i1}$, as shown in Fig. 3(c) and (d). In the configuration of $V_{Sr}$-$2H_{i2}$, two hydrogen atoms are located at the vertexes of two adjacent $SrO_6$ octahedra, with the corresponding O-H bonds oriented towards $V_{Sr}$. Meanwhile, the two hydrogen atoms in $V_{Sr}$-$2H_{i3}$ are located at the vertexes of two next-nearest $SrO_6$ octahedra, with the corresponding O-H bonds oriented towards $V_{Sr}$ too. The lengths of four O-H bonds in these two configurations are all around 0.968 Å as listed in Table I, which is slightly shorter than 0.984 Å reported previously.[19] Moreover, it is evident that the difference in the vibrational frequencies for $b_1$ and $b_2$ in $V_{Sr}$-$2H_{i2}$ and $V_{Sr}$-$2H_{i3}$ is relatively large (about 30 cm$^{-1}$) as shown in Fig. 3(c) and (d). This is different from the previous report, wherein two degenerate modes are obtained.[19] As listed in Table I, the relatively large difference in the vibrational frequencies for $b_1$ and $b_2$ in $V_{Sr}$-$2H_{i2}$ and $V_{Sr}$-$2H_{i3}$ results mainly from the difference in their harmonic frequencies, since the anharmonic contributions are identical.

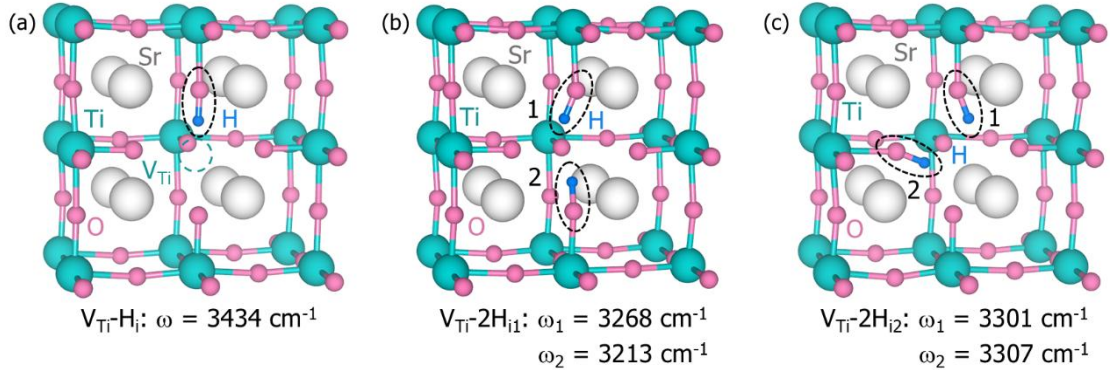

FIG. 4. (a) The configuration of $V_{Ti}$-$H_i$ in -3 charge state along with the calculated vibrational frequency for the corresponding O-H bond; (b) and (c) are the different configurations of $V_{Ti}$-$2H_i$ in -2 charge state along with the calculated vibrational frequencies for the corresponding O-H bonds. 1 and 2 represent the bond indices.

D．Vibrational frequencies of titanium vacancy complexes with interstitial hydrogen

Although titanium atom has a smaller atomic size than strontium atom, titanium vacancy can also trap several hydrogen atoms, forming the $V_{Ti}$-$nH_i$ complexes. For example, the previous study indicates that the titanium vacancy can even trap four hydrogen atoms, forming $V_{Ti}$-$4H_i$ complex.[18] Similar to $V_{Sr}$-$nH_i$, $V_{Ti}$-$H_i$ and $V_{Ti}$-$2H_i$ represent the most important two types of $V_{Ti}$-$nH_i$ complexes as reported in previous studies.[18, 21] Hence, we also focus on these two complexes for $V_{Ti}$-$nH_i$. Fig. 4(a) shows the stable configuration of $V_{Ti}$-$H_i$ in -3 charge state after relaxation, similar to that reported by Varley et al.[21] The O-H bond in this configuration is also oriented towards the $V_{Ti}$, akin to the orientation of O-H bond in $V_{Sr}$-$H_i$. The length of this O-H bond is 0.973 Å, which is very close to 0.975 Å reported by Varley et al.[21] The calculated vibrational properties show that the vibrational frequency of this O-H bond is 3434 cm$^{-1}$, resulting from a harmonic contribution 3633 cm$^{-1}$ and an anharmonic contribution -199 cm$^{-1}$.

For $V_{Ti}$-2$H_i$, our results show that it has the configuration as illustrated in Fig. 4(b) (denoted as $V_{Ti}$-2$H_{i1}$). Compared with the previous report,[21] the orientation of O-H bond $b_2$ in this configuration is slightly different. In the previously reported configuration, the two O-H bonds are nearly centrosymmetric about the $V_{Ti}$, whereas our configuration deviates from this symmetry. In addition, the distribution of bond lengths is also different. The previous work indicates that the two O-H bonds in $V_{Ti}$-2$H_i$ has the same bond lengths 0.975 Å, while our results show that the bond lengths are slightly different, i.e., 0.976 and 0.978 Å as listed in Table I. The discrepancy in bond lengths may result from the different configurations. As shown in Fig. 4(b), the calculated vibrational frequencies of $b_1$ and $b_2$ exhibit a relatively large difference of 55 cm$^{-1}$. This difference is comparable to that (41 cm$^{-1}$) observed in the previous study.[21] Moreover, Limpijumnong et al. proposed another important configuration for $V_{Ti}$-2$H_i$ as shown in Fig. 4(c).[18] In this configuration, the bond lengths of two O-H bonds are identical, i.e., 0.974 Å as listed in Table I. Meanwhile, the calculated vibrational frequencies of these two O-H bonds are also very similar, with 3301 cm$^{-1}$ for $b_1$ and 3307 cm$^{-1}$ for $b_2$. The difference in vibrational frequencies between these two O-H bonds is 6 cm$^{-1}$, which is considerably smaller than 26 cm$^{-1}$ reported by Limpijumnong et al.[18]

In summary, Table I lists all the vibrational properties of O-H bonds in different configurations. As observed in Table I, the vibrational frequency and the bond length basically exhibit an inverse proportional relationship. Specifically, the longer (shorter) bond lengths correspond to lower (higher) vibrational frequencies. Meanwhile, the bond length and vibrational frequency are nearly linear-correlated with each other. This trend is consistent with previous reports,[19,31] thereby validating the robustness of our results. Moreover, it is worth noting that the O-H bonds tend to be oriented towards cation vacancies ($V_{Sr}$ and $V_{Ti}$) as shown in Fig. 3 and 4. This tendency could mainly result from the Coulomb attraction between the negative charged cation vacancies and the positive charged interstitial hydrogen.

### E．Identification of the hydrogen-related absorption bands

In experiment, many studies have performed the measurement of hydrogen-related absorption bands as mentioned in Sec. I. Among these studies, the most comprehensive investigation was conducted by Tarun et al. in 2011.[8] They not only observed the main absorption bands around 3500 cm$^{-1}$, but also discovered the new absorption bands around 3300 cm$^{-1}$. In order to facilitate the identification of these absorption bands, the absorbance data is extracted as shown in Fig. 5. The main absorption bands contain three peaks labeled as $H_I$, $H_{II}$ and $H_{III}$, which are correlated with $H_i$ previously.[17,21] However, based on the appropriate exchange correlation functional, the vibrational frequency of $H_i$ is calculated to be 3277 cm$^{-1}$, which is considerably lower than the previously reported values of 3500 or 3533 cm$^{-1}$. Therefore, our results indicate that $H_i$ is unlikely to be the source of the main absorption bands. Based on the calculated vibrational properties presented in Table I, the strontium vacancy complexes with interstitial hydrogen exhibit the vibrational frequencies around 3500 cm$^{-1}$, and thus should be associated with the main absorption bands. Given that $V_{Sr}$-$H_i$ and $V_{Sr}$-2$H_{i1}$ display vibrational frequencies closer to 3500 cm$^{-1}$, we propose that these defect complexes are the likely sources of the main absorption bands H$_I$, H$_{II}$ and H$_{III}$. This deduction is consistent with the previous theoretical prediction.[19]

For the two new discovered absorption bands labeled as H$_{IV}$ and H$_V$, they are around 3300 cm$^{-1}$ as shown in Fig. 5. The previous report has suggested that these bands might originate from 2H$_i$,[17] while the calculated vibrational frequencies for this configuration are only about 3100 cm$^{-1}$ as shown in Fig. 2(b), which is significantly lower than 3300 cm$^{-1}$. Therefore, 2H$_i$ is unlikely to account for the two new absorption bands. Our results show that V$_{Ti}$-2H$_{i2}$ has the vibrational frequencies around 3300 cm$^{-1}$, suggesting that it could be the source of H$_{IV}$ and H$_V$. Although Limpijumnong et al. also predicted a similar attribution,[18] their calculated vibrational frequencies are only about 3100 cm$^{-1}$ for the same configuration. We attribute the discrepancy primarily to the different used exchange correlation functionals. The accurate prediction of the vibrational frequencies further indicates that our computational method is appropriate and the results are robust. Moreover, Tarun et al. reported the existence of "hidden" hydrogen atoms in STO, which cannot be detected using infrared spectroscopy.[8] Varley et al. have provided a well explanation for these "hidden" hydrogen atoms.[21] Since they cannot be associated with the specific vibrational frequencies, we are not going to discuss them further in this work.

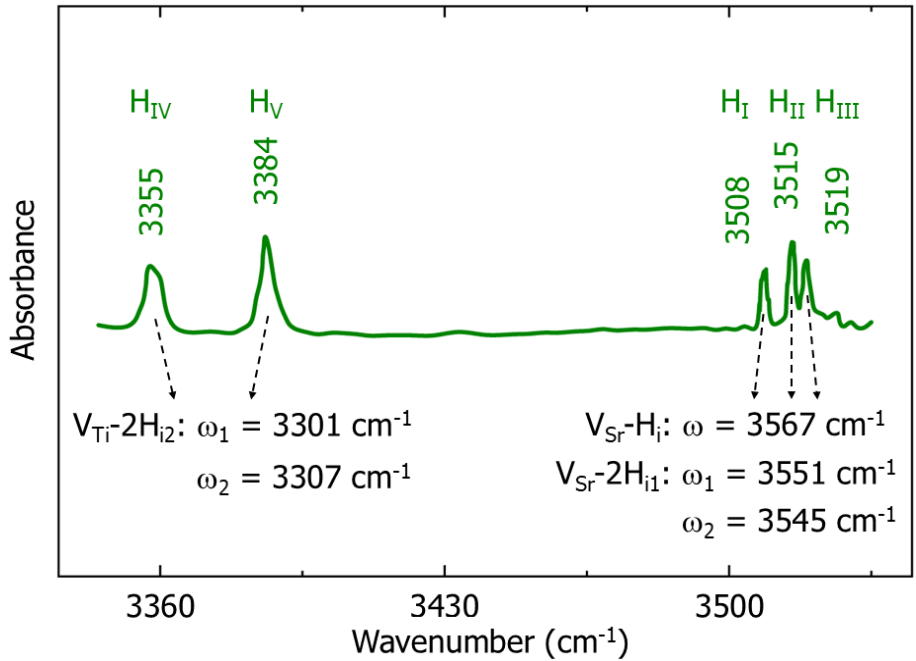

FIG. 5. The proposed configurations of hydrogen impurities corresponding to the experimental absorption bands. The experimental absorbance data is extracted from Ref. 8.

## IV. CONCLUSION

A systematic first-principles study has been performed to identify the configurations of hydrogen impurities in STO. The calculated vibrational frequencies reveal that the interstitial hydrogen (H$_i$) cannot account for the main absorption bands observed experimentally. Instead, our results indicate that strontium vacancy complexes with interstitial hydrogen ($V_{Sr}$-$H_i$ and $V_{Sr}$-$2H_{i1}$) serve as the primary sources of the absorption bands near 3500 cm$^{-1}$, while titanium vacancy complex ($V_{Ti}$-$2H_{i2}$) is likely responsible for the additional bands around 3300 cm$^{-1}$. These results reconcile previous theoretical discrepancies and validate the importance of adopting an appropriate computational approach for investigating the vibrational properties. Overall, this work provides insights into the role of hydrogen-related complexes in governing the electronic properties of STO and advances our understanding of hydrogen-induced defect configurations and in oxide semiconductors.

## ACKNOWLEDGMENTS
This work was supported by National Natural Science Foundation of China (NSFC) under grant No. 12304110.